# Single-hole spin dephasing in bulk crystals.

Yuri A. Serebrennikov

Qubit Technology Center

2152 Merokee Dr., Merrick, NY 11566

ys455@columbia.edu

Decoherence of the hole angular momentum in bulk crystals is described within the framework of non-Markovian stochastic theory. We present the analytical result for the rate of this process. The derivation is based on the new formulation of the multiband Luttinger-Kohn Hamiltonian that reflects the *dynamic* symmetry of the system.

72.25.Rb  03.65.Vf  03.67.Lx

The formidable challenge for spin-based quantum computing lies in the unavoidable coupling of the spin degrees of freedom with the environment that destroys the coherence in a spin subsystem that is vital for quantum logical operations. For carriers possessing nonzero angular momentum the spin-orbit interaction provides the main coupling between the spin and spatial variables and is usually responsible for the fast spin-lattice relaxation process. For instance, holes in bulk semiconductors relax their spin on the scale of the momentum relaxation time and are frequently considered as "spin-relaxed" because of the strong mixing between the spin and orbital degrees of freedom, and the direct coupling of the resultant total angular momentum of the hole with the lattice momentum in the near-degenerate valence bands[1].

The rapid progress of polarization- and time-resolved femtosecond spectroscopy allows the direct measurements of these extremely short spin relaxation times. Recently it has been shown[2] that at room temperature in undoped bulk GaAs relaxation of heavy holes (HH) occurs on the order of 0.1 ps. These results are of paramount importance for the assessment of the feasibility of novel hole-based spintronic devices[3]. Surprisingly, in spite of the large number of theoretical studies of single-hole spin relaxation in nanostructures[1], there does not seem to be any theory of the process in bulk crystals.

Here through the compact procedure that elucidates the underlining physics and correctly reflects the *dynamic* symmetry of the system we will reformulate the $\vec{k} \cdot \vec{p}$ perturbation theory and express the multiband hole Hamiltonian[4] in terms of the irreducible tensor operators of the full rotation group. The primary motivation for this new formalism is the need for an adequate description of the anisotropic part of the *instantaneous* $\vec{k} \cdot \vec{p}$ Hamiltonian, which due to thermal wandering of the lattice momentum $\vec{k}$ through the zone, fluctuates in magnitude and direction, and is zero only under averaging over time. These fluctuations may induce interband transitions and lead to dephasing of the hole angular momentum. The rate of this process will be calculated analytically within the framework of non-Markovian stochastic theory that describes the evolution of the total angular momentum, $\vec{J}$, of the strongly coupled spin-orbit subsystem and provides a clear qualitative picture of the phenomenon.

In bulk semiconductors such as Si, Ge or GaAs in the absence of spin the top of the valence band at the $\Gamma$ point ($\vec{k} = 0$) consist of three degenerate *p*-like functions that transform under the full rotation group as eigenfunctions of the orbital momentum $L = 1$.

In the absence of magnetic fields, near the zone center the second order (effective-mass) $\vec{k} \cdot \vec{p}$ perturbation theory[4,5] gives the following spinless Hamiltonian

$$H_{k^2}^{(L)} = Ak^2 - (A-B)\sum_i k_i^2 L_i^2 - (C/2) \sum_{i,j(i \neq j)} k_i (L_i L_j + L_j L_i) k_j, \quad (1)$$

where $(L)$ denotes the space-fixed frame; $i, j = x, y, z$ represent the Cartesian basis, $\vec{L}$ is the effective orbital angular momentum operator; $A$, $B$, and $C$ are the real constants ("effective masses") defined in Ref.[4]. Within the "spherical approximation"[5,6] cubic terms in the Hamiltonian are neglected, $A$-$B$ is set equal to $C$, and Eq.(1) takes the form

$$H_{k^2}^{(L)} = Ak^2 - (A-B)(\vec{k}\vec{L})^2 . \quad (2)$$

It is easy to see that with the help of the symmetric traceless second rank tensor operator[7] $\vec{Q}_{ij} = [L_i L_j + L_j L_i - (4/3)\delta_{ij}\hat{I}]/2$, where $\hat{I}$ is the identity matrix ($3 \times 3$ in this case), expression (2) can be rewritten as

$$H_{k^2}^{(L)} = (A+2B)k^2/3 - (A-B)\sum_{i,j} k_i \vec{Q}_{ij} k_j . \quad (3)$$

The "quadrupole" Cartesian tensor $\vec{Q}_{ij}$ has five linear independent components that can be expressed in terms of the components of a second rank irreducible tensor operator of the full rotation group[7] $T_{2q}(L) = \sum_{\mu\mu_1} C^{2q}_{1\mu 1\mu_1} L_{1\mu} L_{1\mu_1}$, where $C^{2q}_{1\mu 1\mu_1}$ is the Clebsch-Gordon coefficient. In the *principal-axes* system (*M-frame*) of the $\vec{Q}_{ij}$ tensor[7] Eq.(3) takes the form

$$H_{k^2}^{(M)} = (\gamma_1/2m_0)k^2 + (3\gamma_2/m_0)\{(2/3)^{1/2} D_k T_{20}^{(M)}(1) + E_k [T_{22}^{(M)}(1) + T_{2-2}^{(M)}(1)]\} \quad (4a)$$

or equivalently

$$H_{k^2}^{(M)} = (\gamma_1/2m_0)k^2 + (3\gamma_2/m_0)[D_k (L_{z_M}^2 - 2/3) + E_k (L_{x_M}^2 - L_{y_M}^2)] . \quad (4b)$$

Here we take the hole picture, and reverse the sign of the energy; $m_0$ is the bare electron mass, $\hbar = 1$, the coefficients $\gamma_1$ and $\gamma_2$ are the dimensionless Luttinger parameters[5],

$$D_k := -(2k_{z_M}^2 - k_{x_M}^2 - k_{y_M}^2)/2, \quad E_k := -(k_{x_M}^2 - k_{y_M}^2)/2. \tag{5}$$

Expansion (4a) can be rewritten in the more compact form as

$$H_{k^2}^{(M)} = K + V^{(M)} := K + (3\gamma_2/m_0)\sum_q (-1)^q K_{2q}^{(M)} T_{2-q}^{(M)}(1). \tag{6}$$

Here $K := (\gamma_1/2m_0)k^2$ represents an *isotropic* part of the kinetic energy of the hole. The last term on the RHS of Eq.(6), a scalar product of two spherical tensors of rank-2 ($K_{20}^{(M)} = (2/3)^{1/2} D_k$, $K_{2\pm1}^{(M)} = 0$, $K_{2\pm2}^{(M)} = E_k$), is clearly *anisotropic*, $[L^2, V] \neq 0$. Physically this means that that the ballistic motion of the hole breaks the isotropy of the system and lifts the orbital degeneracy of the *p*-like functions that exists only at the $\Gamma$-point ($\vec{k} = 0$). The anisotropic part of the spinless $\vec{k} \cdot \vec{p}$ Hamiltonian, *V*, reflects the wandering of $\vec{k}$ through the zone and may be zero only under averaging over time. We will return to this point below. Thus, even if the carrier equilibrium distribution in the *k*-space is isotropic, the *instantaneous* Hamiltonian of the crystal outside the zone center lacks spherical symmetry. (This fact was overlooked in Ref.[6].) For axially symmetric systems ($E_k = 0$) $L_{z_M}$ does commute with $H_{k^2}^{(M)}$ and Eq.(4) yields eigenstates that can be characterized by the helicity $m_L = \vec{k} \cdot \vec{L}^{(M)}/|\vec{k}|$; $m_L = \pm 1$, $m_L = 0$ with the eigenvalues $E_{\pm 1} = K + (\gamma_2/m_0)D_k$ and $E_0 = K - (2\gamma_2/m_0)D_k$. In fact, with the replacement $\vec{L} \to \vec{I}$, where $\vec{I}$ is an effective operator of the unit spin, $I^2 = I(I+1) = 2$, and apparent redefinition of the parameters one may recognize in Eq.(4b) the familiar fine structure spin Hamiltonian of the ion in the anisotropic crystal field[8].

The spin of the carrier doubles the degeneracy of the valence bands at the $\Gamma$-point. In the presence of *k*-independent *intrinsic* spin-orbit coupling (SOC)[9], $H_{SO} = \lambda \vec{L} \cdot \vec{S}$, the Bloch functions are not factorizable into the orbital and spin parts, hence, the total angular momentum is required to characterize the basis kets. The combined action of the *isotropic* SOC and the *anisotropic V* will split the sixfold degenerate manifold of the valence bands into a series of states. For a given value of $\vec{k}$ the corresponding basis can be built from the spherical spinor functions of the compound *L-S* system[7]

$$|LS, Jm; \vec{k}> = \sum_{\mu,\mu_1} C^{Jm}_{L\mu\, 1/2\,\mu_1} |L\mu; \vec{k}> |1/2\,\mu_1>.$$ The matrix elements of $H_{SO}$ and $T_{2q}(L)$ in this basis are well known[7], which allows to represent the $6 \times 6$ matrix ($L = 1$, $S = \frac{1}{2}$) of the $\vec{k} \cdot \vec{p}$ Hamiltonian, $H^{(M)} = H_{SO} + H^{(M)}_{k^2}$, in the following form

$$<1\,1/2, J_1 m_1; \vec{k}|H^{(M)}|1\,1/2, Jm; \vec{k}> = \frac{\lambda}{2}[J(J+1) - \frac{11}{4}]\delta_{JJ_1}\delta_{mm_1} + K\delta_{JJ_1}\delta_{mm_1}$$
$$+ (-1)^{J+3/2}[5(2J+1)]^{1/2}(3\gamma_2/m_0)\begin{Bmatrix} 2 & 1 & 1 \\ 1/2 & J_1 & J \end{Bmatrix}[\sqrt{\frac{2}{3}}D_k C^{J_1 m_1}_{Jm\,20} + E_k(C^{J_1 m_1}_{Jm\,22} + C^{J_1 m_1}_{Jm\,2-2})] \quad (7)$$

By inspection, it is easy to see that Eq.(7) recovers the Luttinger-Kohn Hamiltonian[4] (*in the M-frame*). Notably, the $J = J_1 = 3/2$, $4 \times 4$ block of this matrix represents the Luttinger Hamiltonian[5] (columns below correspond to m = 3/2, ½, -1/2, -3/2):

$$<1\,1/2, 3/2 m_1; \vec{k}|H^{(M)}|1\,1/2, 3/2 m; \vec{k}> = \frac{\gamma_2}{m_0}\begin{pmatrix} D_k & 0 & \sqrt{3}E_k & 0 \\ 0 & -D_k & 0 & \sqrt{3}E_k \\ \sqrt{3}E_k & 0 & -D_k & 0 \\ 0 & \sqrt{3}E_k & 0 & D_k \end{pmatrix} \quad (8)$$

Note that, $[J^2, H^{(M)}] \neq 0$, however, if *V* is axially symmetric ($E_k = 0$), then $J_{z_M}$ is conserved and the eigenfunctions of $H^{(M)}$ can be classified by the helicity $m = \hat{\vec{k}} \cdot \vec{J}^{(M)}$.

Bands with $J = 3/2$, $m = \pm 3/2$ correspond to the HHs; bands with $J = 3/2$, $m = \pm 1/2$ represent the light holes (LHs). Finally, bands with $J = 1/2$, $m = \pm 1/2$ correspond to the split-off (SO) holes. Due to the *T*-invariance of the problem (no magnetic interactions) each of these bands has Kramers degeneracy.

Thus, similar to a fine structure splitting in isolated atoms, SOC breaks up the six-fold valence band degeneracy at $\Gamma$-point into multiplets of $\vec{J}$ ($\Gamma_8$ and $\Gamma_7$, splitting $= 3\lambda/2$), but preserves the isotropy of the system. Anisotropy comes from the ballistic motion of the hole that shifts the carrier from the center of the zone and, similar to a crystal field, is responsible for further lifting of the degeneracy of the $\Gamma_8$ states into HH and LH bands. The intrinsic SOC is rather strong in common semiconductors, e.g., $3\lambda/2 = 340$ mEv in GaAs and it is usually safe to ignore mixing between $J = 3/2$ and $J = 1/2$ bands. In this approximation the eigenvalues of the axially symmetric $6 \times 6$ Luttinger-Kohn Hamiltonian Eq.(7) are $E_{HH} = \lambda/2 + K + (\gamma_2/m_0)D_k$ and $E_{LH} = \lambda/2 + K - (\gamma_2/m_0)D_k$, $E_{SO} = -\lambda + K$.

The main advantage of the expansion (6) is the simplicity of the transformation of irreducible tensor operators under rotations of the coordinate system[7]

$$V^{(L)}(\Omega_t) = D^{-1}(\Omega_t) V^{(M)} D(\Omega_t) = \frac{3\gamma_2}{m_0} \sum_{qp} (-1)^p T_{2p}^{(L)}(1) D^2_{q,-p}(\Omega_t) K_{2q}^{(M)}. \quad (9)$$

Here $D(\Omega_t)$ is the operator of finite rotation, the set of Euler angles $\Omega_t$ represents the instantaneous orientation of the laboratory (*L*) frame relative to the *M*-frame of reference at the moment *t*, $D^2_{q,-p}(\Omega_t)$ is the corresponding Wigner rotation matrix. With the help of Eq.(9), the $6\times 6$ Luttinger-Kohn Hamiltonian Eq.(7) can be represented in the *L*-frame in the following compact form

$$H^{(L)}(t) = H_0 + V^{(L)}(\Omega_t) := \lambda \vec{L}\cdot\vec{S} + \frac{\gamma_1}{2m_0}k^2 + \frac{3\gamma_2}{m_0}\sum_{qp}(-1)^p T_{2p}^{(L)}(1) D^2_{q,-p}(\Omega_t) K_{2q}^{(M)}. \quad (10)$$

Thermal motion of $\vec{k}$ in the crystal results in the random modulation of $V$. The latter connects the *L-S* subsystem to the bath and is responsible for interband transitions and $\vec{J}$-dephasing process. Expression (10) provides the foundation for the theoretical study of this process presented below.

It is customary to treat the problem of spin relaxation in the Markovian limit. This approach is based on the assumption that correlation times of the bath are negligibly short compared to the characteristic timescale of the spin subsystem. Consequently, the spin decoherence does not depend on the evolution of the compound system (spin + bath) over times comparable with or less than the randomization time $\tau_c$ of those degrees of freedom of the bath that participate in the interaction ($\sim 10^{-12}-10^{-13} s$ in the condensed phase). Markovian approximation is well justified in systems with zero (quenched) orbital angular momentum and large gap between electron states of different orbital symmetry, where SOC is suppressed and the residual spin-lattice interactions, e.g., magnetic dipole-dipole or hyperfine are relatively weak (~ tens MHz). However, this may not be the case for hole (or electron) states with nonzero orbital angular momentum. In such systems, spin and orbital degrees of freedom are strongly coupled and the equation of motion of the full *L-S* + *V* system must be solved *before* the calculation of any macroscopic observable. As already mentioned, strong SOC in common semiconductors leads to the large splitting of $\Gamma_8$ and $\Gamma_7$ valence bands with the frequency of the corresponding coherent motion of $\vec{J}$ in the range of 10 - 100 THz. At this timescale, lattice phonons involved in the "collisions" with carrier have not enough scattering events to establish thermal equilibrium. Moreover, these frequencies of motion in the strongly coupled *L-S* subsystem are comparable with or even higher than the characteristic frequencies of coherent motion of the vibrational excitation in solids and it is necessary to consider a finite duration of a multiparticle interaction (phonon picture may be not adequate). As a result, during the transient period, $t \sim \tau_c$, that takes reservoir to randomize $\vec{k}$, environment retains "memory", $\vec{J}$ is entangled with $\vec{k}$, and the evolution of the hole angular momentum must be described within the framework of non-Markovian kinetics.

In the "fast motional" limit, i.e., when the *action* of anisotropic part of the Luttinger-Kohn Hamiltonian is weak,

$$<V^2>\tau_c^2 \sim (4\gamma_2^2/m_0^2)<D_k^2>\tau_c^2 <<1 \quad , \tag{11}$$

the standard Fano-Zwanzig projection operator technique[10] leads to the following kinetic equation for the ensemble averaged, $<...>$, operator of the total angular momentum

$$<\dot{\vec{J}}(t)> = -\int_0^t \hat{M}(t-t')<\vec{J}(t')>dt' , \tag{12}$$

where $\hat{M}(\tau)$ is the memory superoperator. The first nonvanishing term in the cumulant expansion of $\hat{M}(\tau)$ yields[10]

$$\hat{M}(\tau) = <V^\times(\tau)\exp(iH_0^\times \tau)V^\times(0)> . \tag{13}$$

Here $A^\times a := [A,a]$ and it is assumed that due to the spherical symmetry of the problem the hole equilibrium distribution in the *k*-space is isotropic, $<V(t)> = 0$ and $H_0^\times <\vec{J}(t)> = 0$.

The open upper limit of the time integration in Eq.(12) reflects the non-Markovian character of the theory. Note, that in the "coarse-graining" Markovian limit the relaxation operator does not depend on time and, hence, contains no information about the evolution of the system over times $t \leq \tau_c$. The autocorrelation function $K_{\vec{J}}(t) := Tr[\rho_J^{eq} \vec{J}^+ \vec{J}(t) >]$, where $\rho_J^{eq}$ is the equilibrium density operator can be found with the help of Laplace transformation, $\widetilde{X}(p) := \int_0^\infty \widetilde{X}(t) \exp(-pt) dt$:

$$\widetilde{K}_{\vec{J}}(p) = Tr\{\rho_J^{eq} \vec{J}^+ [p\hat{1} + \hat{\widetilde{M}}(p)]^{-1} \vec{J}\}. \tag{14}$$

To proceed further it is convenient to choose the orthonormal set of polarization operators[7] $T_{kq}(J) = \sum_{\mu \mu'} C_{J\mu\,kq}^{J\mu'} | LS, J\mu' >< LS, J\mu | := | T_{kq}(J) >>$ as the basis of the corresponding Liouville space. Substitution of Eqs.(9), (10), and (13) into Eq.(14) yield after summation over magnetic quantum numbers the following non-model result (for simplicity we assume that $V$ is axially symmetric)

$$\widetilde{K}_{\vec{J}}(p) = (1/3) \sum_{q, J, J_1} \rho_J^{eq} \Pi_{JJ_1} << T_{1q}(J) | [p\hat{1} + \hat{\widetilde{M}}(p)]^{-1} | T_{1q}(J_1) >>, \tag{15}$$

$$\hat{\widetilde{M}}_{JJ_1}(p) = \frac{4\gamma_2^2}{m_0^2} <D_k^2> \begin{bmatrix} 2\widetilde{\Phi}_2(p)/5 + 1/2 \operatorname{Re} \widetilde{\Phi}_2(p + i\omega_0) & 1/\sqrt{10} \operatorname{Re} \widetilde{\Phi}_2(p + i\omega_0) \\ 1/\sqrt{10} \operatorname{Re} \widetilde{\Phi}_2(p - i\omega_0) & \operatorname{Re} \widetilde{\Phi}_2(p - i\omega_0) \end{bmatrix} \tag{16}$$

Here the first and second columns represent bands with $J = 3/2$ and $J = 1/2$; $\omega_0 := 3\lambda/2$, $\Pi_{JJ_1} := [J(J+1)(2J+1)J_1(J_1+1)(2J_1+1)]^{1/2}$, and $\widetilde{\Phi}_2(p + i\omega_0)$ is the spectral density of the normalized to unity autocorrelation function

$$\Phi_2(t) = 5 < D_k(t) D_{00}^2[\Omega_t] D_k(0) D_{00}^2[\Omega] > / <D_k^2> . \tag{17}$$

By inspection of Eqs.(15) and (16), one may conclude that in the fast-motional limit $V$ is self-averaged by rapid isotropic fluctuations of the orientation of $\vec{k}$ and the spherical symmetry of the system is restored; $J$ is a good quantum number. Although, similar to the crystal filed splitting, anisotropic part of the instantaneous Hamiltonian Eq.(10) will split bands with $J > 1/2$ differing in $|m_J|$, one cannot distinguish in Eq.(16) between the HH, $m_{J=3/2} = \mp 3/2$, and the LH, $m_{J=3/2} = \pm 1/2$ components of the $\Gamma_8$ quadruplet. In this regime thermal fluctuations of $\vec{k}$ lead to virtual transitions within the $J = 3/2$ quadruplet, proportional to $\tilde{\Phi}_2(p)$, and may invoke nonadiabatic "jumps" between $\Gamma_8$ and $\Gamma_7$ valence bands. The latter are described by the elements of $\hat{\tilde{M}}_{JJ_1}(p)$ that are proportional to $\text{Re}\tilde{\Phi}_2(p \pm i\omega_0)$. Notably due to the fundamental symmetry restriction, $V$ does not split and induce transitions between the components of the pure SO Kramers doublet, $m_{J=1/2} = \pm 1/2$.

The value that is of main interest is the dephasing time $T_{\tilde{j}} := \tilde{K}_{\tilde{j}}(p=0)/K_{\tilde{j}}(0)$, which generally depends on the explicit form of the autocorrelation function $\Phi_2(t)$, strength of $V$, and the correlation time $\tau_c = \tilde{\Phi}_2(0)$. If fluctuations of $V$ are treated as an external Markovian noise (semiclassical approximation, bath at infinite temperature), then $\Phi_2(t) = \exp(-t/\tau_c)$, $\text{Re}\tilde{\Phi}_2(i\omega_0) = \text{Re}\tilde{\Phi}_2(-i\omega_0) = \tau_c/[1+(\tau_c\omega_0)^2]$ and it is easy to see from Eq.(16) the *quadratic* decrease of the probability of interband transitions with the growth of SOC in the system, if the adiabatic condition $\omega_0\tau_c \gg 1$ is fulfilled.

This prediction, however, is unacceptable since in accordance with adiabatic theorem one should expect an exponential decline of the rate of transitions between the widely separated electronic states with the growth of $\lambda$. On the other hand, the non-Markovian stochastic theory, even in semiclassical approximation, may lead to the correct functional dependency of the spectral density upon the level spacing in a quantum subsystem. To illustrate this, consider, for example, the following model function $\Phi_2(t) = (2\tau_c/\pi)^2/[t^2 + (2\tau_c/\pi)^2]$, which yields $\operatorname{Re}\tilde{\Phi}_2(i\omega_0) = \tau_c \exp(-2\tau_c\omega_0/\pi)$. Thus, the account for the short time, $t \sim \tau_c$, evolution of the $L$-$S + V$ system within the framework of non-Markovian kinetics allows the proper description of the adiabatic (i.e., exponential) reduction of the probability of interband transitions with the growth of SOC.

If we completely neglect the nonadiabatic transitions the matrix Eq.(16) becomes diagonal and it is easy to see that in the limit of strong SOC the adiabatic fluctuations of $V$ ($\omega_0 \tau_c \gg 1$) do not lead to dephasing of the pseudospin-1/2 that represents SO band; $1/T_{\vec{J}}(\Gamma_7) \to \infty$. The same adiabatic perturbation, however, may scramble entanglement inside the $\Gamma_8$ quadruplet. It follows from Eqs.(15) and (16) that the rate of this process is described by the expression that does not depend on the explicit form of $\Phi_2(t)$:

$$1/T_{\vec{J}}(\Gamma_8) = (8\gamma_2^2/5m_0^2) < D_k^2 > \tau_c . \qquad (17)$$

Note that the rate of $\vec{J}$-dephasing is smaller than $1/\tau_c$, similar to Dyakonov-Perel mechanism of spin relaxation[1] $T_{\vec{J}}$ is inversely proportional to $\tau_c$. Although $T_{\vec{J}}(\Gamma_8)$ depends upon the parameters $< D_k^2 >$ and $\tau_c$ that are difficult to evaluate, Eq.(17) may be used to obtain reasonable estimates of the magnitude of the effect. If we assume that at the room temperature $2\sqrt{< D_k^2 >} \sim < k^2 > \approx 10^{-3}/a_0^2$, where $a_0$ is the lattice constant (see

Ref.[11]), then for GaAs $(2\gamma_2/m_0)\sqrt{<D_k^2>} \approx 40$ mEv. For $\tau_c = 80$ fs this gives $T_{\bar{j}} \approx 0.3$ ps that has the correct order of magnitude[2].

The applicability of Eq.(17) is restricted by the condition of the "fast motional" approximation, Eq.(11), which requires unrealistically short $\tau_c$ to wash out an average LH-HH splitting larger than 40 mEv. In fact, the *distinct* optical orientation and relaxation of HHs and LHs that was clearly observed in Ref.[2] cannot be described by the theory presented here. Nevertheless, our non-model results suggest that unresolved HH and LH bands (closer to the zone center) may be consistent with relatively long $T_{\bar{j}}$. The absence of the hole optical orientation does not necessarily mean that the rates of angular and linear momentum relaxation are the same.